\documentclass[]{emulateapj}
\usepackage{natbib}

\shorttitle{[OIII] emission from globular clusters NGC~4472}
\shortauthors{Peacock et al.}

\begin{document}

\title{Limits on [OIII]~5007 emission from NGC~4472's globular clusters: constraints on planetary nebulae and ultraluminous black hole X-ray binaries in globular clusters}
\author{Mark B. Peacock and Stephen E. Zepf}
\affil{Department of Physics and Astronomy, Michigan State University, East Lansing, MI 48824, USA}
\email{MBP: mpeacock@msu.edu}
\and
\author{Thomas J. Maccarone}
\affil{School of Physics and Astronomy, University of Southampton, Southampton, SO17 1BJ, UK}

\begin{abstract}
\label{sec:abstract}

We have searched for [OIII]~5007 emission in high resolution spectroscopic data from Flames/Giraffe VLT observations of 174 massive globular clusters (GCs) in NGC~4472. No planetary nebulae (PNe) are observed in these clusters, constraining the number of PNe per bolometric luminosity, $\alpha<0.8\times10^{-7}$PN/L$_{\odot}$. This is significantly lower than the rate predicted from stellar evolution, if all stars produce PNe. Comparing our results to populations of PNe in galaxies, we find most galaxies have a higher $\alpha$ than these GCs (more PNe per bolometric luminosity -- though some massive early-type galaxies do have similarly low $\alpha$). The low $\alpha$ required in these GCs suggests that the number of PNe per bolometric luminosity does not increase strongly with decreasing mass or metallicity of the stellar population. We find no evidence for correlations between the presence of known GC PNe and either the presence of low mass X-ray binaries (LMXBs) or the stellar interaction rates in the GCs. This, and the low $\alpha$ observed, suggests that the formation of PNe may not be enhanced in tight binary systems. These data do identify one [OIII] emission feature, this is the (previously published) broad [OIII] emission from the cluster RZ~2109. This emission is thought to originate from the LMXB in this cluster, which is accreting at super-Eddington rates. The absence of any similar [OIII] emission from the other clusters favors the hypothesis that this source is a black hole LMXB, rather than a neutron star LMXB with significant geometric beaming of its X-ray emission.

\end{abstract}


\section{Introduction}
\label{sec:intro}

Emission lines in the spectra of extragalactic globular clusters (GCs) provide a useful method for identifying exotic objects in their integrated light. In particular, the presence of the [OIII]~5007 emission line in a GC is indicative of the presence of objects such as planetary nebulae (PNe), supernova remnants or X-ray binaries. Despite our interest in identifying and studying such objects, and the frequency with which they occur, few systematic searches for emission lines in the spectra of GCs currently exist. 

One of the main sources of [OIII] emission in a stellar population are PNe. These nebulae are produced near the end of an asymptotic giant branch (AGB) star's life when gas, driven outward by a wind or superwind, is ionized by the hot inner core. Our understanding of the formation of these systems remains uncertain. In particular, challenges remain in explaining the axisymmetric shape of some PNe \citep[e.g.][]{Soker06,DeMarco09}, the universality of the brightest PNe observed across a wide range of galaxy morphologies \citep[e.g.][]{Ciardullo02,Ciardullo10} and the number density of PNe in different galaxies \citep{Hui93,Ciardullo05,Buzzoni06}. In recent years, theories for the formation of PNe have suggested that stellar interactions in binary systems may play an important role \citep[e.g.][]{Moe06,Soker06,DeMarco09}. The population of PNe formed in GCs provides an interesting addition to studies of PNe. GCs provide a relatively well constrained stellar population in which to study PNe in old, metal poor, environments. They also provide a unique dynamical environment in which to consider the effects of binaries and stellar interactions on the formation of PNe. 

In the Galactic GC system, searches for PNe have identified only four such systems in 133 clusters \citep{Jacoby97}. Based on the total mass of these clusters, \citet{Jacoby97} proposed that PNe may be under abundant in globular clusters, compared with galactic fields (although at a significance of only around 3$\sigma$). In extragalactic GC systems, PNe can be identified in their unresolved spectra via the strong [OIII] emission from these nebulae. However, there have been few studies dedicated to identifying emission lines from GCs and currently only a few extragalactic GC PNe have been recorded. These were identified via the 5007$\rm{\AA}$ emission line from GCs in NGC~3379 \citep{Bergond06,Pierce06}, the Fornax dwarf \citep{Larsen08} and NGC~7457 \citep[which may be an intermediate age cluster,][]{Chomiuk08}. In this paper, we examine high resolution spectra of GCs around NGC~4472, with the aim of investigating the formation of PNe in GCs. These clusters have a combined luminosity of $\sim$2$\times10^{8}$L$_{\rm{\odot,V}}$, $8\times$ more than the Galactic GC system. 

Another potential source of [OIII] emission from GCs are nebulae produced around bright X-ray sources. X-ray and [OIII] emission has previously been observed from a GC in NGC~4472 \citep{Zepf07} and a GC in NGC~1399 \citep{Irwin10}. The [OIII] emission observed from the NGC~4472 GC is very broad, implying velocities of $\sim$2000~km/s \citep{Zepf08}. This is too high to be from a PN in the cluster. Instead it is likely that this emission is associated with the bright black hole low mass X-ray binary (LMXB) in the cluster. This X-ray binary is thought to have super-Eddington accretion rates and is likely ionizing a wind driven from the system \citep{Maccarone07,Zepf07}. The detection of similar (broad) [OIII] emission in GCs provides a method for identifying more of these interesting systems and investigating the properties of X-ray binaries with very high accretion rates.

\section{NGC~4472 Data/ reductions}
\label{sec:data}

This study is based on archival spectra of GCs in the galaxy NGC~4472. The survey targeted known clusters from the catalog of \citep{Rhode01} in four different fields around NGC~4472. Details of each observed field are given in table \ref{tab:fields}. These data were obtained on various nights between May to July 2003 and January to March 2006, under the programs 072.B-0384(A) and 075.B-0516(A) (PI: Bergond). Here, we review these data, but further details of this and similar surveys can be found in \citet{Bergond06,Zepf07} and Zepf et al. (in prep). 

These data were obtained on the very large telescope (VLT) using the FLAMES instrument in its GIRAFFE/MEDUSA mode \citep{Pasquini02}. This is a fiber fed, multi-object spectrograph, consisting of 130 fibers which can be positioned over a 25$^{\prime}$ field. The spectra cover the range 5010$< \lambda < $5810$\rm{\AA}$ with 0.2$\rm{\AA}$~pixels and a resolution of 0.9$\rm{\AA}$.

We obtained the raw science data from the ESO archive\footnote{http://archive.eso.org} and reduced it using the \textsc{GASGANO} software\footnote{http://www.eso.org/sci/software/gasgano/}. Master dark and bias frames were produced for each observation using the routines gimasterdark and gimasterbias. In the absence of the original reduced calibration files, we used the standard calibration data sets provided by the GIRAFFE pipeline (as recommended in the pipeline documentation). These include the flatfielding/ transmission correction, fiber location/ width determination and dispersion solution. The data were then reduced using the giscience routine in  GASGANO. This routine performs the bias, dark and flatfield corrections. It then applies the dispersion solution and extracts each spectrum from the frame. The spectra were then associated with the targets using the fiber association data table (included in the ESO archive data). 

The sky was measured independently for each exposure via dedicated sky fibers, spread across the field of view. The sky level was not found to correlate strongly with either location in the field of view or distance to the center of NGC~4472. The lack of a correlation with galactocentric radius is unsurprising. This is because GCs in the central regions of NGC~4472 were not targeted by this survey, therefore the galaxy surface brightness is relatively faint and fainter than the sky for most regions. For the most central GC, which is $\sim3\arcmin$ from the center of NGC~4472, we expect the galaxy to contribute $\sim$25$\%$ of the GC flux. Given the lack of correlations, the improved S/N produced by a simple average of all sky fibers was found to produce the best results. The cluster spectra were sky subtracted and combined using the \textsc{IRAF} tasks \textsc{IMARITH/SCOMBINE}. An approximate flux calibration was then applied to each cluster spectrum based on its known V-band magnitude from \citet{Rhode01}. 

This provided 453 spectra of NGC~4472's GCs. However, several clusters were observed multiple times (in different fields). In total, these data provide spectra for 358 unique clusters. Where multiple spectra were available for a cluster, these were averaged together. 

The GCs in galaxies like NGC~4472 cover a wide range of radial velocities. For example, \citet{Cote03} show that NGC~4472's GCs have velocities in the range $250< v < 1950$~km/s, corresponding to wavelength shifts of $4 < \Delta\lambda < 32\rm{\AA}$. It is therefore difficult to identify 5007$\rm{\AA}$ emission from clusters without determining their redshift. As part of a separate study, these data have been used to derive the radial velocities of 174 of these clusters (Zepf et al., in prep.). For the remaining clusters, the S/N ratio was too low to obtain reliable velocities. The high resolution of these data allowed velocities to be estimated with errors $<$20~km/s. We used these velocities to estimate the rest frame wavelengths of these data with $\Delta\lambda<0.3\rm{\AA}$. 

\begin{table}
 {\centering
  \caption{Targetted fields \label{tab:fields}}
  \begin{tabular}{@{}cccccc@{}}
   \hline
   \hline
   Field &   RA$^{1}$  & DEC$^{1}$ & Number & Total exp & Number          \\
         &                  &                  & exp      & time (ks)  & clusters$^{2}$ \\
   \hline
   NE   &  187.56317 & 8.10417  &       4         &      10.0         &    108 ( 80)     \\
   SE   &  187.55866 & 7.88373   &       6         &      15.6         &    114 (101)    \\
   SW  &  187.33344 & 7.88362   &       5         &      11.5         &    112 ( 74)    \\
   NW  &  187.35649 & 8.14778  &       4         &      10.4         &    117 ( 87)     \\
   \hline
   \end{tabular}\\
 }
\footnotesize
$^{1}$approximate center of each field. \\
$^{2}$parentheses indicates the number of clusters for which their velocities can be measured from these data (Zepf et al., in prep). \\
\end{table}

\section{5007$\rm{\AA}$ emission from NGC~4472's GCs} 
\label{sec:results}

\begin{figure*}
 \centering
 \includegraphics[height=170mm,angle=270]{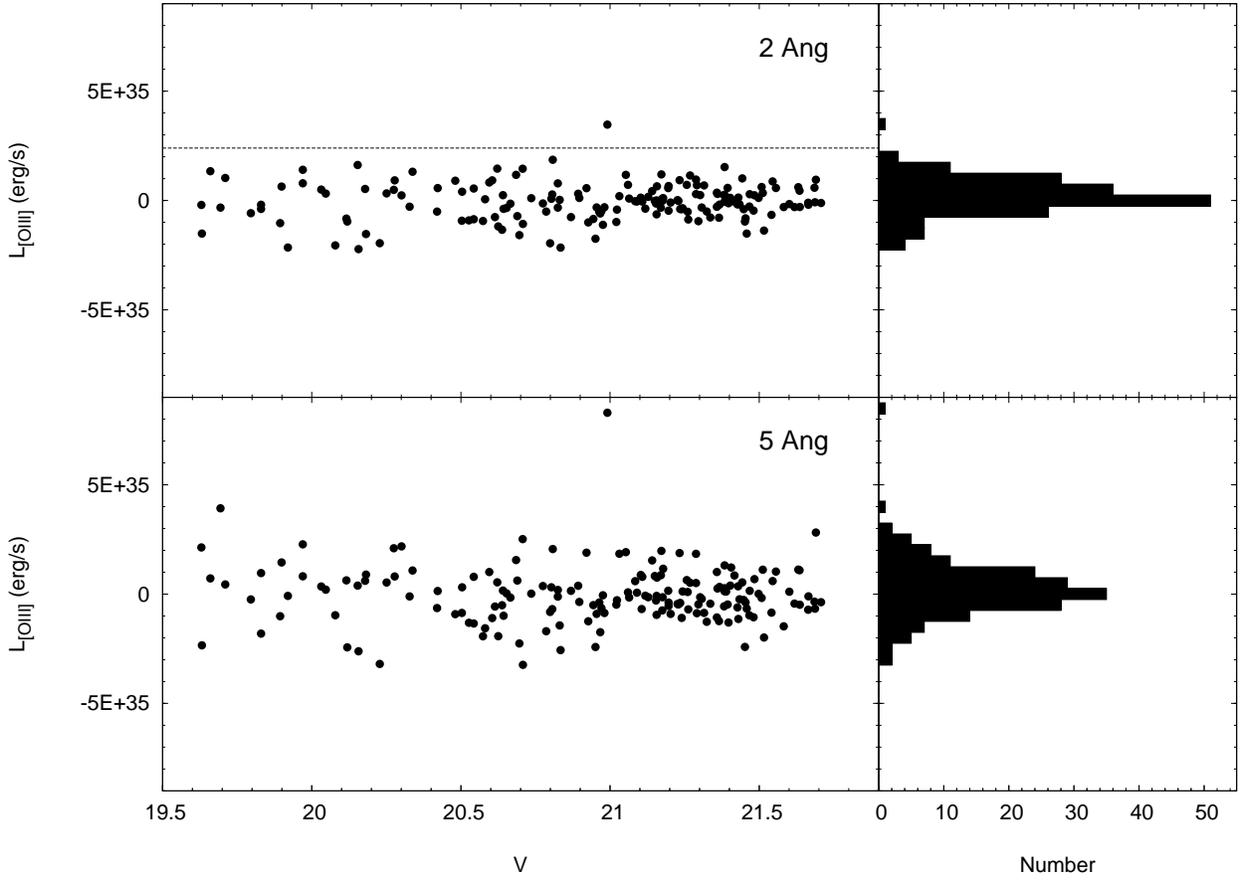} 
 \caption{NGC 4472 GC luminosities in the [OIII]5007 emission line as measured
with respect to the continuum, in a 2$\rm{\AA}$ (top) and 5$\rm{\AA}$ (bottom) region centered on rest-frame 5007$\rm{\AA}$. Positive luminosities indicate emission features above the continuum, while negative luminosities represent absorption features. The dashed line corresponds to the luminosity of a PN 2.5~mag beneath the brightest PN cutoff luminosity. The right hand panels show histograms of the measured luminosities. They demonstrate that the measured luminosities are consistent with Gaussian noise distributed around zero. One cluster can be identified in these data as having excess emission over these regions. This is the (previously identified) cluster RZ~2109. It can be seen that the luminosity of this emission feature increases significantly in the wider line region, suggesting a very broad emission feature. }
 \label{fig:L_oiii} 
\end{figure*}

We used these data to identify 5007$\rm{\AA}$ emission from the 174 GCs with known velocities. This was done by measuring the flux at the (redshift corrected) wavelength of 5007$\rm{\AA}$ using two boxcars with width 2 and 5$\rm{\AA}$. For each cluster, we fit the continuum level via a linear fit to the spectrum, from 5010-5060$\rm{\AA}$, excluding the region 4$\rm{\AA}$ either side of the line. The 2$\rm{\AA}$ width measurement is appropriate for measuring emission from PNe, which typically have velocities of $\sim$50km/s. Therefore the large majority of their flux should be in this narrow region. The wider (5$\rm{\AA}$) bin has greater scatter, due to the increased noise associated with the wider region, but is more sensitive to broad emission features from non-PNe sources, such as that seen from RZ~2109.

Figure \ref{fig:L_oiii} shows the 5007$\rm{\AA}$ luminosities of these clusters. In this figure, we present line strengths as luminosities, rather than equivalent widths. This is because the clusters in our sample have different luminosities (and hence different continuum levels). The line luminosities are relative to the continuum level, such that emission and absorption lines have positive and negative fluxes, respectively. The right panels of figure \ref{fig:L_oiii} show the distribution of 5007$\rm{\AA}$ line luminosities. It can be seen that the measured line luminosities have an approximately Gaussian shape centered close to zero luminosity and with a standard deviation of 8.5$\times10^{34}$erg/s/$\rm{\AA}$. This is similar to that expected from the noise of the average reduced cluster spectra. These have $\sigma \simeq 6 \times 10^{34}$erg/s/$\rm{\AA}$, and thus $\sigma$ in our 2$\rm{\AA}$ bins is $\sqrt{2}$ times this
value, or $\sigma(2\rm{\AA}) = 8.5\times10^{34}$erg/s/$\rm{\AA}$. Such a distribution would be expected due to noise, if no 5007$\rm{\AA}$ features are present in the clusters. We therefore assume that the scatter observed in figure \ref{fig:L_oiii} is due primarily to noise. Studies of PNe often quote the population of PNe within 2.5~mag of the brightest PN cutoff magnitude (M*), where (${\rm M}^{*}_{\rm{[OIII]}}$-2.5) = -1.98 (corresponding to a luminosity, L$_{\rm{[OIII]}}\sim2.4\times10^{35}$erg/s). This limit is plotted in figure \ref{fig:L_oiii} as the dashed line. It can be seen that PNe above this limit would be detected by these data. 

The top panel of figure \ref{fig:L_oiii} identifies only one cluster with emission above this limit. The luminosity of this cluster increases significantly in the wider 5${\rm \AA}$ region, suggesting that this emission is very broad. This emission has previously been identified in this cluster \citep[RZ~2109,][]{Zepf07} and was studied in greater detail by \citet{Zepf08}, who identified [OIII]~5007 and [OIII]~4959 lines with widths of $\sim$2000km/s. No clusters with similar, broad, emission are identified in these data. In section \ref{sec:BH-LMXBs}, we discus this result in the context of the population of ultraluminous LMXBs in GCs. 

No other emission features are observed in these data (we confirmed our automated approach by visually examining the cluster spectra for emission lines). This suggests that no PNe, in the brightest 2.5~mag of the planetary nebula luminosity function (PNLF), are present in these clusters. The implications of this non-detection on the GC PN population is discussed in the following section.

\section{Planetary nebulae in globular clusters}
\label{sec:PN}

\subsection{How many PNe are expected?} 
\label{sec:alpha}

The absence of any PNe in these data raises the question of how many are expected. To investigate this, we consider some simple theoretical predictions for the number of PNe in a stellar population, per bolometric solar luminosity, $\alpha$\footnote{We use $\alpha$ to represent the total population of PNe and $\alpha_{x}$ to represent the population with [OIII] magnitudes in the range M* to M*-$x$}. An estimate for $\alpha$ can be obtained from simple stellar evolution to be, $\alpha\sim2\times10^{-11}\tau_{\rm{PN}}\rm{PN/L}_{\odot}$ \citep[][]{Renzini86,Jacoby97,Buzzoni06}. This is the number of post AGB stars formed (per solar luminosity) $\times$ the lifetime of an observable PN, $\tau_{\rm{PN}}$. While the lifetime of PNe may vary in different stellar populations, we can estimate an upper limit as the dynamical timescale, $\tau_{\rm{PN,dyn}}\sim30,000~\rm{yr}$. This is the time it takes the nebula gas to disperse (and hence end the observable PN phase). These considerations suggest that, if all stars produce PNe with lifetimes close to the dynamical timescale, $\alpha\sim6\times10^{-7}\rm{PN/L}_{\odot}$. Some stellar populations may have shorter PNe lifetimes, or may not produce observable PNe at all, therefore this value is an estimate for the maximum number of PNe expected, per solar luminosity. 

\subsubsection{$\tau_{{\rm PN}}$ and age}
\label{sec:alpha-age}

The age of a stellar population can have a significant effect on $\alpha$. \citet{Buzzoni06} considered the effect, on $\alpha$, of the decreasing mass of PNe central stars, with the increasing age of the stellar population. They demonstrated that initially $\tau_{\rm{PN}}$ (and hence $\alpha$) should increase, with increasing age, due to the slower evolution (and hence longer lifetimes) of the lower mass central stars. However, as the age increases further, it takes the central stars longer to reach the temperatures required to ionize the PN, hence reducing their observable lifetime (and reducing $\alpha$). Eventually, one would expect the central star mass to fall beneath a critical limit of $\sim$0.52$\rm{M_{\odot}}$. At these low masses, the star will evolve too slowly to reach the temperatures required to ionize the circumstellar gas, and hence produce an observable nebula. This limit may be important for very old stellar populations, such as GCs and the oldest populations in galaxies. \citet{Moe06} investigated the number of PNe expected in a stellar population with an age of 11.5~Gyr and a turnoff mass of 0.85M$_{\odot}$ -typical of a GC. Their conclusion was that no PNe are likely to be produced in such a population, based on the evolution of isolated stars. However, they note that there is an uncertainty in this prediction, due to the uncertainty in the central star low mass cutoff (and the relationship between main sequence turnoff mass and the resulting PN central star mass). 

\subsubsection{Binary interactions}
\label{sec:alpha-binary}

The above estimates for $\alpha$ consider the evolution of an isolated star. However, it is likely that binary interactions play an important role in producing PNe. It has been proposed that most (if not all) of the PNe that are observed are produced through some form of binary interactions \citep[e.g.][]{Bond00,DeMarco04,Soker06,DeMarco06,DeMarco09}. There are a few reasons to suspect that binary interactions are important in the formation of PNe. Firstly, the brightest PNe (M*) observed in galaxies across all morphologies are found to be very similar, regardless of the age of the stellar population. This is not expected from the evolution of a single star. For example, the simulations of \citet{Marigo04} suggest that M* should decrease by approximately 3~mag in the first 6~Gyr of star formation \citep{Ciardullo10}. Secondly, most PNe observed are found to be non-spherical. Both of these observational constraints are hard to produce from the evolution of a single star, but can potentially be explained in binary scenarios \citep[see e.g.][]{Ciardullo05,Soker06,DeMarco09}. If binary interactions are required to produce observable PNe, then we require that the PN predecessor needs to be in a binary system which is tight enough that interactions can occur during the AGB phase. Additionally, the binary separation must be large enough that interactions do not occur at earlier times, such as during the red giant branch phase, when the core is not hot enough to ionize a nebula. \citet{DeMarco05} propose that, this limits PN forming systems to binaries with separations ($a$) of roughly $100<a<500\rm{R_{\odot}}$. In a later paper, \citet{DeMarco09} estimate that 13$\%$ of stars reside in binaries in which such interactions can occur, with a corresponding reduction in $\alpha$. The binary formation mechanism therefore predicts significantly lower $\alpha$ than would be expected based solely on stellar evolution. 

\subsubsection{$\alpha$ in galaxies}
\label{sec:alpha-galaxies}

Empirical estimates for $\alpha$ are also available for several nearby galaxies. However, it should be noted that, obtaining accurate measurements for $\alpha$ is challenging. In the Milky Way, estimates for $\alpha$ vary significantly, partially due to difficulties in measuring the distances to Galactic PNe. For example, \citet{Phillips02} estimated $\alpha\sim6.3\times10^{-7}$ in the Milky Way, while \citet{Soker06} suggested that the distances to the PNe considered by \citet{Phillips02} may be, on average, 1.5$\times$ greater, implying $\alpha\sim1.9\times10^{-7}$. The best constrained PNe populations are those of the SMC and LMC. For these galaxies large numbers of PNe (with known distances) can be detected across a significant fraction of the PNLF. For more distant galaxies, beyond our local group, PNe surveys are often constrained to the brightest tip of the PNLF (with detection limits ranging from 0.5-2.5~mag fainter than M$^{\star}$). Their total population of PNe is then extrapolated, based on an assumed PNLF. \citet{Buzzoni06} collated and presented $\alpha$ for both local group galaxies and a sample of early type galaxies (many of which we have plotted in figure \ref{fig:alpha}). This work, and the references therein, demonstrates that there are significant variations in $\alpha$. While some galaxies have $\alpha$ in the range of the upper limit, discussed in section \ref{sec:alpha}, others are a factor $\sim$7 lower. Despite these variations, most galaxies are found to have $\alpha$ in the range of $1-6\times10^{-7}$PN/L$_{\odot}$ \citep[e.g.][]{Jacoby93,Buzzoni06,Ciardullo10}. 

Studies of the PN population in galaxies have observed correlations between a galaxy's properties and its population of PNe. Although the reason is yet to be fully understood, it is observed that more massive, metal rich galaxies tend to have have lower $\alpha$ \citep[e.g.][]{Peimbert90,Hui93,Ciardullo05,Buzzoni06}. Another correlation is observed between $\alpha$ and the far-ultraviolet (FUV) excess of a galaxy (its FUV-V color). FUV bright galaxies appear to produce less PNe per solar mass \citep[e.g.][]{Jacoby93,Ciardullo05,Buzzoni06}. Such an effect may be expected. The FUV excess in early type galaxies is now thought to be due to extreme horizontal branch (EHB, or subdwarf~B) stars \citep[e.g.][]{Dorman95,OConnell99}. Such stars may skip the AGB phase of stellar evolution entirely and proceed directly to the white dwarf phase as AGB-manqu\'{e} stars \citep{Greggio90}. Therefore, stellar populations which produce relatively more EHB stars are likely to be under abundant in PNe. It should be noted that the masses, metallicities and ages of early type galaxies are related -- with massive, metal rich galaxies having older stellar populations \citep[e.g.][]{Trager00}. Therefore, any/ all of these paramters could drive the observed PN relationship. 

\subsection{Known PNe in globular clusters} 
\label{sec:PNGC}

\begin{table*}
 {\centering
  \scriptsize
  \caption{Planetary nebulae in globular clusters \label{tab:GCPN}}
  \begin{tabular}{@{}crcccrcrccl@{}}
   \hline
   \hline
    Names     & (m-M)$^{1}$ & M$_{\rm{V}}^{1}$ & [Fe/H]$^{1}$ & log$(\Gamma_{c})^{1,2}$ & F$_{5007}$    & m$_{5007}$ &  L$_{5007}$ & M$_{5007}$ & M$^{\star}$-M & references \\
  host~galaxy/GC/PN    &           &                       &    &      & erg/s/cm$^{2}$ &                &  $\times$10$^{35}$erg/s   &                &                &                 \\
   \hline
  MW/M15/Ps1-K648 & 15.09 & -9.19 & -2.37 & 6.83  & 1.87E-12 & 15.58 & 0.242 & 0.49 & 4.97 & \citet{Adams84} \\
  MW/M22/GJJC1       & 12.53 & -8.50 & -1.70 & 5.63  & 3.0E-13 & 17.57 & 0.004 & 5.04 & 9.52 & \citet{Gillett89} \\
  MW/Pal6/JaFu1        & 13.82 & -6.79 & -0.91 & 5.28  & 1.25E-10 & 11.02 & 5.032 & -2.8 & 1.68 & \citet{Jacoby97} \\
  MW/NGC6441/JaFu2 & 15.32 & -9.63 & -0.46 & 7.17  & 1.9E-13 & 18.06 & 0.031 & 2.74 & 7.22 & \citet{Jacoby97} \\
  NGC3379/gc771/PN & 29.91 & -8.65 & -1.40 &  - & 1.54E-17 & 28.29 & 1.700$^{3}$ & -1.62 & 2.86 & B06, P06, P11$^{3}$\\
  Fornax-dSph/H5/PN & 20.68 & -7.40 & -1.73 &  - & 5.0E-15 & 22.01 & 0.112 & 1.33 & 5.81 & \citet{Larsen08} \\
  NGC7457/GC7/PN    & 30.55 & -8.20 & -0.40 &  - & 6.04E-17 & 26.81 & 12.000 & -3.74 & 0.74 & \citet{Chomiuk08} \\
   \hline
   \end{tabular}\\
 }
\footnotesize
$^{1}$for the Galactic GCs, this data was obtained from the Harris catalog \citep[2010 edition;][]{Harris96} \\
$^{2}$the stellar interaction rate in the cluster core, $\Gamma_{c} \propto \rho^{3/2}/r^{2}$ \citep[see e.g.][]{Verbunt87} \\
$^{3}$this source was identified by both \citet[][B06]{Bergond06} and \citet[][P06]{Pierce06}. However, the [OIII] luminosity is not quoted in these papers. We measured this from the Flames/Giraffe spectrum used by \citet{Bergond06}. This spectrum was obtained and reduced in a similar fashion to the NGC~4472 data, discussed here, and the line measured using IRAF/SPLOT. \\
\end{table*}

\begin{figure*}
 \centering
 \includegraphics[width=170mm,angle=0]{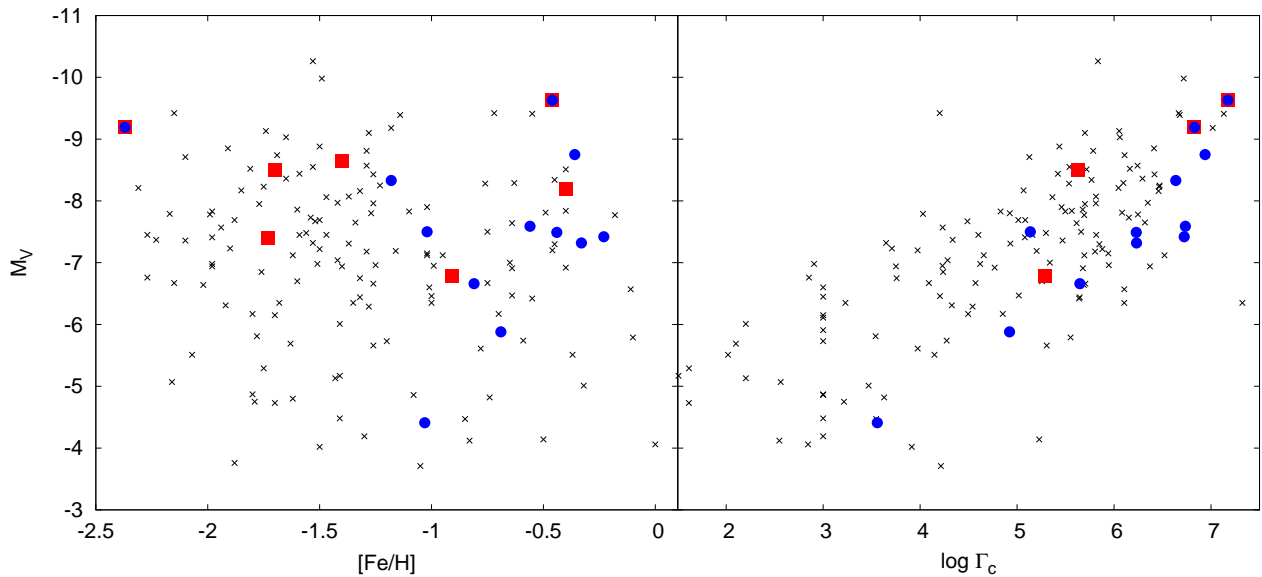} 
 \caption{Absolute V-band magnitude of Milky Way GCs (crosses) as a function of [Fe/H] and stellar interaction rate in the cluster core, $\Gamma_{c}$. The PNe hosting clusters are highlighted as red squares. For the magnitude and metallicity, we show all known GC PN, but $\Gamma_{c}$ is only available for the four Galactic PN GCs. For comparison, the LMXB hosting clusters are highlighted as blue circles. }
 \label{fig:gcpn} 
\end{figure*}

In table \ref{tab:GCPN}, we list all of the GC PNe that are currently proposed. In the Galactic GC system only four PNe have been found \citep{Jacoby97}. For extragalactic GC systems relatively few PNe are known, partially due to a lack of data and studies aimed at identifying emission lines in GCs. Three PNe have been proposed in extragalactic GCs in: the Fornax~dSph, \citet{Larsen08}; NGC~3379, \citet{Bergond06}, \citet{Pierce06}; and NGC~7457, \citet{Chomiuk08}. In addition to these GC PNe, [OIII] emission has been observed from several other clusters. One cluster in M~31, appears to have [OIII] emission, which may be due to a PN, but this is yet to be studied \citep[as can be seen in the spectra of the GC B115;][]{Kim07,Caldwell11}. Emission observed from several other extragalactic GCs may not be be due to PNe. These include two clusters in M87 \citep[where the OIII emission is thought to originate from background filaments in the central region of M87;][]{Cohen98}, a cluster in NGC~5128 \citep[which was identified as the first PN in an extragalactic GC, but higher resolution spectroscopy has subsequently shown that the OIII emission is double peaked, suggesting velocities which exclude a PN as the origin;][]{Minniti02,Peng04} and a second source in NGC~7457 \citep[with an uncertain origin, but possibly from a supernova remnant;][]{Chomiuk08}. 

Table \ref{tab:GCPN} lists the properties of these PNe and the clusters which host them. It can be seen that the PNe span a wide range of [OIII] luminosities, from close to the tip of the PNLF to 9~mag fainter. Because of the low number of GC PNe known, and the varying survey depths, the GC PNLF is hard to constrain from this sample. However, there is no evidence for either brighter or fainter PNe being favored in GCs, relative to Galactic regions. It is also interesting to consider the properties of the clusters which host a PN, since this may provide clues to the origin of GC PNe. The left panel of figure \ref{fig:gcpn}, shows the metallicity and luminosity of the Galactic GCs (crosses) and of the GCs which host PNe (red squares). As discussed in section \ref{sec:alpha-galaxies}, a correlation is observed between $\alpha$ and the metallicity of early type galaxies, with metal poor galaxies hosting more PNe. It can be seen that no metallicity effects are observed in these GCs. A K-S test confirms that the metallicity of the PN hosting GCs is consistent with being drawn randomly from the Galactic GC population. The PN hosting clusters do favor more massive clusters. This may simply be a consequence of the greater number of stars in these clusters. \citet{Jacoby97} also noted that 2/4 of the Galactic GCs, which host a PN, also host an LMXB. From this observation, they suggest that PNe and LMXBs may form preferentially in the same clusters (albeit with a significance $<3\sigma$). Such a correlation would be interesting, since the presence of an LMXB in a cluster is considered indicative of the formation of tight binary systems. Therefore, this may suggest a possible binary origin for the GC PNe. A more direct way to study such a relationship is by considering the stellar interaction rate in the dense cores of these clusters, $\Gamma_{c} \propto \rho^{3/2}/r^{2}$ \citep[where $\rho_{c}$ and $r_{c}$ are the cluster's core density and radius;][]{Verbunt87}. If the formation of PNe is related to binary interactions, then they should be found primarily in clusters with high $\Gamma_{c}$. A strong correlation is observed between $\Gamma_{c}$ and the formation of LMXBs in the GC systems of the MW \citep[e.g.][]{Verbunt87}, NGC~5128 \citep{Jordan07} and M31 \citep{Peacock10b}. The right panel of figure \ref{fig:gcpn} shows the stellar interaction rate for all Galactic GCs and those which host PNe (red squares) and LMXBs (blue circles). Both the LMXB and PN hosting clusters have higher than average $\Gamma_{c}$. However, due to the low number of PNe, the difference between the Galactic GCs and the PN GCs is \textit{not} significant. Furthermore, it can be seen that there is a strong correlation between the luminosity of a cluster and its interaction rate. While the LMXB hosting clusters have significantly higher than average $\Gamma_{c}$ for their luminosity, strongly supporting a dynamical formation origin, the PN GCs do not. 

From the known GC PNe, we therefore conclude that there is no evidence for enhanced dynamical formation of PNe in these GCs. The partial correlation between the presence of a PN and an LMXB, proposed by \citet{Jacoby97}, is likely a consequence of the high stellar mass in the two clusters which host both a PN and an LMXB. The PNe are found preferentially in more massive clusters. This is consistent with their formation from either isolated stars or primordial (rather than dynamically enhanced) binary interactions. 

\subsection{$\alpha$ in globular clusters}
\label{sec:PNGC-alpha}

\begin{figure*}
 \centering
 \includegraphics[width=170mm,angle=0]{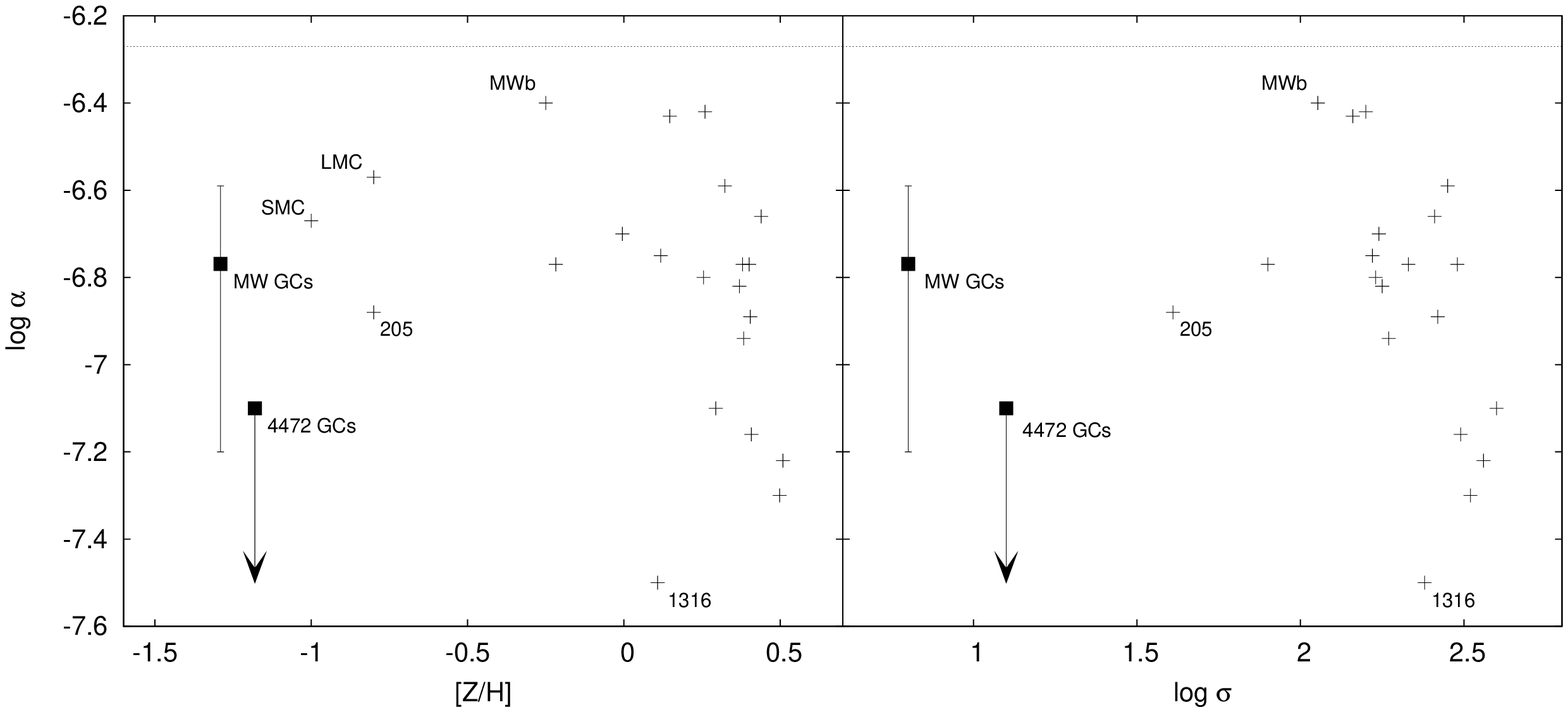} 
 \caption{Number of PNe per bolometric luminosity, $\alpha$, as a function of the host stellar population's metallicity ([Z/H], left) and velocity dispersion ($\sigma$, right). Crosses show $\alpha$ for a sample of early type galaxies, the MW bulge, LMC and SMC \citep[taken from tables 4 and 6 of][]{Buzzoni06}. Filled squares show the MW's and NGC~4472's GC systems. See text for details on the data used. The dashed line is the predicted $\alpha$ assuming all AGB stars produce PNe with with lifetimes close to the dynamical timescale of the nebula. It can be seen that the observed $\alpha$ in these galaxies increases with decreasing mass and metallicity. However, this correlation does not appear to extend to the lower mass and metallicity GC systems. }
 \label{fig:alpha} 
\end{figure*}

Due to the low number of GC PNe currently known, and the lack of systematic searches for such objects in large GC samples, $\alpha$ in GCs is currently poorly constrained. An empirical estimate for $\alpha$ can be produced based on the Galactic GCs, although this estimate suffers from small number statistics. \citet{Jacoby97} identified four PNe in Galactic GCs with a combined stellar luminosity of 2.4$\times10^{7}$L$_{\odot}$, implying $\alpha$ of 1.7$\times10^{-7}$PN/L$_{\odot}$. This survey covered the brightest $\sim$9~mag of the PNLF, significantly deeper than most extragalactic surveys of PNe can achieve. For surveys sensitive to the brightest 2.5~mags of the PNLF (such as the data considered in this paper), only one PN, that in Pal~6, would be detected in the Galactic GCs (see table \ref{tab:GCPN}). It should also be noted that the association of this PN, with Pal~6, is relatively uncertain \citep[with a 0.5$\%$ chance that they are uncorrelated,][]{Jacoby97}. The absence of any PNe in the NGC~4472 cluster data provides further important constraints on the formation of PNe in GCs. The combined stellar luminosity of these clusters is $\sim$2$\times10^{8}$L$_{\rm{\odot,V}}$ ($\sim$8$\times$ that of the Galactic clusters). This corresponds to a bolometric luminosity, L$_{\rm{bol}}\sim$2.8$\times10^{8}$L$_{\odot}$ \citep[assuming L$_{\rm{bol}}$ for an old simple stellar population is $\sim$1.4$\times$L$_{\rm{V}}$,][]{Maraston05}. The detection of no PNe in these GCs therefore corresponds to $\alpha_{2.5}\le$0.8$\times10^{-8}$PN/L$_{\odot}$ \citep[where the upper limit is based on the 90$\%$ confidence level for a non-detection,][]{Kraft91}. This is smaller than that measured for the similar stellar populations of the Galactic GCs. However, the numbers are consistent, given the low number of sources detected. 

Both GC systems host significantly less PNe than potentially possible for their stellar mass, see section \ref{sec:alpha}. \citet{Jacoby97} have previously demonstrated that the Galactic GCs host less PNe than this limit, with a significance of 3.1$\sigma$. The non-detection of any PNe in the larger stellar mass covered by NGC~4472's GCs strengthens this finding, with $\alpha\lesssim 7 \times$ lower than predicted. This suggests that, either some of the AGB stars in GCs do not evolve into PNe, or the lifetimes of PNe in GCs are significantly shorter than the dynamical timescale. The lower $\alpha$ in these GCs could be due to their old stellar populations having PN central star masses that are too low to produce observable PNe. However, given the simple stellar populations and similar (old) ages of the GCs, this raises the further question of how the PNe that \textit{are} observed in the Galactic GCs were produced. One possible explanation is that the GC PNe are the evolved stage of blue straggler stars. Blue stragglers are known to be present in GCs and can have masses significantly greater than the main sequence turnoff. If only blue stragglers are capable of producing PNe in GCs then their rarity may explain the lower rates observed. \citet{Ciardullo05} have previously discussed the possible role of blue stragglers in producing the brightest PNe observed in galaxies.  Unfortunately, the low number of PNe in the Galactic GCs makes a direct comparison with their blue straggler population impossible. However, observations of the Galactic GC PNe lends some support to this theory. The masses of the central stars in two of these PNe have been measured at 0.62 and 0.75M$_{\odot}$ \citep{Alves00,Harrington93}. Such stars would have main sequence masses much higher than the main sequence turnoff mass in these clusters. 

As discussed in section \ref{sec:alpha-binary}, binary interactions may be required to produce PNe, with only 13$\%$ of stars thought to have binary companions in the required range. Therefore, if binary interactions are required to produce PNe, one would expect a corresponding reduction in $\alpha$. The resulting rate would be in reasonable agreement with both the MW's GCs and the upper limit imposed by NGC~4472's GCs.  However, this does not explain the lower fraction of PNe in GCs compared with some galactic populations. A potentially relevant observation is that primordial binary fractions may be lower in the Galactic GCs than in the Galactic field \citep[although such fractions are very difficult to determine; e.g.][]{Moni_Bidin09}. If binary interactions are required, then differing binary fractions/ parameters may correspond to the different $\alpha$s observed. 

It should be noted that, binary systems are significantly effected by the dense stellar environments in GCs. New binary systems can be formed via tidal capture \citep{Fabian75} or direct collisions \citep{Verbunt87} and binaries are driven to shorter periods via dynamical \citep{Heggie75} and exchange \citep{Hills76} interactions. The result is that GCs contain more close binaries, per stellar mass, than galactic field regions \citep[as confirmed via the dramatic excess of accreting compact objects in GCs, e.g.][]{Clark75}. Therefore, if binary interactions are important in producing PNe, one may expect this process to be enhanced in GCs. This does {\it not} appear to be the case in these GCs, which have $\alpha$ lower than that of many galaxies. This suggests that, close interacting binaries may not enhance the formation of PNe. In fact, by driving binaries to shorter periods, it may be that binary interactions reduce the number of PNe in GCs. If the binaries have shorter periods, and hence their stars interact in the early AGB phase, or even during the RGB phase, the envelopes of these stars may be ejected from the system before the central star is hot enough to ionize the gas and hence produce a PN. Interactions with a companion star on the RGB have previously been proposed to enhance the formation of EHB stars in dense GCs \citep{Buonanno97, Han07, Peacock10c}. Since EHB stars may skip the AGB phase entirely, such interactions may reduce the number of PNe in these clusters. A similar argument may be related to the low $\alpha$ of FUV bright ellipticals, which are thought to have higher fractions of EHB stars. 

Figure \ref{fig:alpha} compares $\alpha$ for the MW's GC system, NGC~4472's GC system and a sample of galaxies \citep[taken from tables 4 and 6 of][]{Buzzoni06}. It can be seen that $\alpha$ for the Galactic GCs \textit{is} consistent with that measured in many early type galaxies. However, there is a large uncertainty in $\alpha$ for the Galactic GCs, due to their low combined stellar mass and the low number of PNe associated with them. The non-detection of any PNe in NGC~4472's clusters requires that these GCs have one of the lowest $\alpha$s of any stellar population observed (only 18$\%$ of the galaxies shown in figure \ref{fig:alpha} have $\alpha$ beneath the upper limit required for these GCs). However, some galaxies \textit{do} have $\alpha$ that is consistent with the limit suggested for NGC~4472's GCs. 

In figure \ref{fig:alpha} we plot $\alpha$ as a function of the metallicity and velocity dispersion ($\sigma$) of the stellar populations. For the early type galaxies their velocity dispersion and Mg$_{2}$ lick indices are taken from table 4 of \citet{Buzzoni06}. We convert Mg$_{2}$ to [Z/H] using stellar population model predictions for a 12Gyr old population, assuming [$\alpha$/H]=0.3 \citep[][]{Thomas03, Maraston05, Thomas11}\footnote{taken from http://research.icg.port.ac.uk/$\sim$thomasd/tmj.html}. For local group galaxies the metallicities were taken from the RGB star observations of \citet[][SMC]{Carrera08b}, \citet[][LMC]{Carrera08a} and \citet[][NGC~205]{McConnachie05}. The metallicity and velocity dispersion of the Galactic bulge were taken from \citet{McWilliam94} and \citet{Minniti96}, respectively. Similarly to individual stars in galaxies, the metallicity of individual GCs in both the MW's and NGC~4472's GC systems varies (with GC values in the range -2.5$\lesssim$[Z/H]$\lesssim$0). We plot the average values of the GCs in both systems. The metallicity and $\sigma$ quoted for the MW GC system is the mean of the GC values, as presented in the catalog of \citet{Harris96}. For NGC~4472's GC system we have converted the GC's mean B-R color and V-band magnitudes \citep[taken from the catalog of][]{Rhode01} to [Z/H] and $\sigma$ using the correlations observed between these parameters in the MW's GCs. The previously observed correlation between $\alpha$ and the mass/ metallicity of the galaxies can be seen. However, these correlations do not seem to extend to the relatively low mass, metal poor, GCs -- which have much lower $\alpha$ than would be predicted by the galaxy relationships. The reason for these correlations in the galaxy data is still uncertain, making it is hard to understand the implications of them breaking down in GCs. If the PN population is effected directly by either metallicity or mass, the lower rates observed in GCs could be due to other reductive processes, such as an age cut off or binary interactions, dominating over these effects. Another potential explanation for the observed rates is that the correlations observed are driven by the ages of the stellar populations. The massive, metal rich, early type galaxies will tend to have older ages, closer to those of the GCs \citep[e.g.][]{Trager00}. Therefore, if the correlations observed in figure \ref{fig:alpha} are due to population age, rather than metallicity or mass directly, one may expect the GCs to have a similar PN population to the massive high metallicity older galaxies, as observed. 

Investigating age effects directly in this sample of GCs is hard due to their similar (old) ages. Our results agree with the low values predicted by \citet{Buzzoni06} for these old populations. \citet{Buzzoni06} also predict that young stellar populations should have low $\alpha$. This is because the young central stars in the PNe evolve quickly, shortening the lifetimes of the PNe. \citet{Larsen06} have previously studied a sample of 80 massive young clusters (with ages around 100~Myrs) in several nearby galaxies. They identified four PNe candidates in these clusters. This suggests that these young populations do have a low $\alpha$ of $0.6\times10^{-7}$PN/L$_{\odot}$ -- similar to the low value found for the MW's and NGC~4472's GCs (with expected ages $>$10~Gyrs). As can be seen from figure 15 of \citet{Buzzoni06}, higher values of $\alpha$ would be expected at intermediate ages of a few Gyrs. Investigating the PN population of such clusters would be highly desirable. Unfortunately, identifying intermediate age clusters has proved very challenging \citep[e.g.][]{Kundu05,Larsen05,Zepf09}.

\section{Black hole LMXBs in globular clusters}
\label{sec:BH-LMXBs}

\subsection{Nebular emission from ultraluminous LMXBs}

In section \ref{sec:results}, we demonstrated that the only cluster in our sample which shows [OIII] emission is, the black hole (BH) LMXB hosting cluster, RZ~2109. This cluster is interesting because it hosts one of the few known BHs in a GC. One of the most conclusive signatures of a BH primary in an LMXB, is to detect variable X-ray emission that greatly exceeds the Eddington limit for an accreting neutron star. Using this method, BHs have been proposed in this GC \citep{Maccarone07}, a second GC in NGC~4472 \citep[not covered by our survey,][]{Maccarone11a}, a GC in NGC~3379 \citep{Brassington10} and in two GCs in NGC~1399 \citep{Shih10,Irwin10}. It is thought that LMXBs accreting at such high rates (at or above the Eddington limit) can drive strong winds from the systems, which will likely be ionized by the central LMXB \citep{Begelman06,Zepf08}. This would suggest that ultraluminous LMXBs may generally be associated with nebula emission and is the proposed source of the [OIII] emission observed in RZ~2109. Unfortunately, the other BH-LMXB candidates cannot be used to test this ubiquity. The [OIII] emission from three of the candidates is currently unknown. The other ultraluminous BH candidate is associated with [OIII] and [NII] emission lines. However, this source has different properties to RZ~2109 and may be an intermediate mass BH accreting from a tidally disrupted while dwarf \citep{Irwin10} or a stellar mass BH ionizing the wind of a R~Corona Borealis star \citep{Maccarone11b}.

\subsection{Ultraluminous LMXBs: BHs or beamed NSs?}

The detection of X-ray emission, with luminosities in excess of the Eddington limit for an accreting NS, provides some of the best evidence for BHs in GCs. However, it has also been proposed that such luminosities can be produced by an accreting NS in an ultracompact X-ray binary, if the X-ray emission is geometrically beamed towards us \citep{King01}. The Eddington limit for accretion on to a 1.4$\rm{M}_{\odot}$ NS is $\sim1.8\times10^{38}$~erg/s. If we assume hydrogen poor accretion, as may be expected from a white dwarf donor in an ultracompact LMXB, slightly higher Eddington limits of $\sim3.5\times10^{38}$~erg/s can be reached. The observed luminosity (L$_{\rm{obs}}$) can then be enhanced, along the line of sight, via geometric beaming, to give L$_{\rm{obs}}$=L$_{\rm{Edd}}/b$. Here, $b$ is the beaming factor (=$\Omega/4\pi$). Hence beaming factors of 0.1-0.3 would be required to produce the observed luminosities of the BH LMXB candidates. Significantly, in such models, the [OIII] emission should not be beamed. This is because the nebula emission has to come from much larger scales than the X-ray emitting region of the binary. 

If [OIII] emission, such as that observed in RZ~2109, is associated with LMXBs accreting at super-Eddington rates, this provides a way to discriminate between beamed-NS and unbeamed-BH models. Under the beaming scenario, there should be, on average, 1/$b$ additional LMXBs accreting at super-Eddington rates, but beamed away from our line of sight. If these systems are present, their (unbeamed) [OIII] emission should be similar to that observed from binaries in which the X-ray emission is beamed towards us. We would therefore expect to observe [OIII] emission from $(1/b)\times N_{\rm{sEdd}}$ clusters, where $N_{\rm{sEdd}}$ is the number of LMXBs with observed super-Eddington accretion rates. From the spectra studied in this paper, we can confirm that there are no additional RZ~2109 like sources in this sample. Based on the observation of one bright LMXB in these data, the above beaming factors would predict an additional 3-10 [OIII] sources. Our non-detection of any sources therefore favors mild beaming factors $>$0.4 (based on a confidence level of 0.90 for the non-detection). It should be noted that, this conclusion assumes that the one super-Eddington LMXB in our sample is representative of the global population. Currently, the number of super-Eddington LMXBs is quite poorly constrained, since we are dealing with very small numbers of sources. Considering this uncertainty, our data suggests that high beaming factors are unlikely, but milder factors of $\sim$0.3, such as those proposed by \citet{King11}, cannot be ruled out. Therefore, we conclude that the non-detection of additional [OIII] sources in these clusters favors the BH LMXB hypothesis with mild or no beaming. We stress that more [OIII] data and new detections of super-Eddington LMXBs should help to better constrain these conclusions in the future. 

Additional to the lack of [OIII] emission in other GCs, there are difficulties in reaching the high X-ray luminosities using the beamed NS LMXB hypothesis. \citet{King11} suggests that the BH candidate in RZ~2109 (XMMU~122939.7+075333), another NGC~4472 GC BH candidate (CXOU~1229410+075744) and the hyperluminous source HLX-1 can be explained as super-Eddington NSs in ultracompact LMXBs. The reason for invoking ultracompact LMXBs is the claim that they can have higher effective Eddington luminosities, so that they need to exceed their Eddington luminosities by smaller factors in order to appear at the observed luminosities.  It is widely discussed that for Type I X-ray bursts, this is the case -- as discussed above, hydrogen-poor material will have an effective Eddington limit almost twice as high as that for hydrogen rich material \citep[e.g.][]{Kuulkers03}. However, this is not the case for ultraluminous X-ray sources, which have characteristic temperatures nearly an order of magnitude lower than those for Type I X-ray bursts.  At $\rm{kT} \sim 0.1-0.2$ keV, as is typically seen for ULXs, the effective Eddington luminosity for material from a carbon-oxygen white dwarf becomes dramatically lower than it is for hydrogen rich material, except at very low densities. This is because the opacity stops being dominated by Thomson scattering, and begins to be dominated by carbon and oxygen edges \citep[see e.g.][]{Iglesias93}. 

If we take parameter values suggested by \citet{King11} for XMMU~122939.7+075333 (in RZ~2109); an accretion rate of $10^{19}$ g/sec and a 1.4$\rm{M}_\odot$ NS. The spherization radius, where the accretion disk solution becomes locally Eddington limited, occurs at $4\times10^7$~cm.  The accretion disk solution should approximate a slim disk not too far outside this radius.  One can estimate the density of the accretion flow as $\dot{M} t_{visc}/(4\pi R_{sp}^3)$, where $\dot{M}$ is the accretion rate into the outer disk, $t_{visc}$ is the viscous timescale from the spherization radius, and $R_{sp}$ is the spherization radius. Setting $t_{visc}$ to $\alpha_{vis}^{-1}(H/R)^{-2} t_\phi$, where $H$ is the scale height of the disk, $\alpha_{vis}$ is the viscosity parameter in standard accretion disk models, $R$ is the radius of the accretion disk, and $t_\phi$ is the orbital period in the accretion disk and taking $(H/R)=0.1$ and $\alpha_{vis}=0.1$ (both typical parameters), we find that the density of the accretion disk will be about $10^{-3} \rm{g/cm}^{3}$. One can then examine figure 9 of \citet{Iglesias93}, and find that the opacity per unit mass will be about 10 times as large for a carbon/oxygen mixture at 2-3 million K than it will be for pure hydrogen gas at the same temperature.  As a result, the effective Eddington limit will be a factor of 10 {\it lower} than the standard Eddington luminosity, rather than a factor of 2 {\it higher} than the standard Eddington luminosity, and both a larger spherization radius, by a factor of about 3 and a larger effect from geometric beaming (by a factor of about 10) are needed relative to the proposal from \citep{King11}.  The accretion disk would then be expected to peak in the extreme ultraviolet, rather than in the X-rays, and the model becomes incapable of explaining the high X-ray luminosities observed.  If we consider a BH, rather than a NS accretor, the density of the gas will be sufficiently low that it will be highly ionized -- enough that the opacity will not be far different from the Thomson opacity. The result is that, for a BH accretor, there is only a minor perturbation to the accretion solution by having a donor made of carbon and oxygen, rather than a donor of roughly solar composition. 

Additionally, for the other BH candidates considered by \citet{King11}, CXOU~1229410+075744 and HLX-1, the model fails to fit their observed X-ray spectra.  In both cases, the observed thermal components are far too hot to be explained by a model involving super-Eddington accretion without the model having much larger beaming factors than those invoked in the model of \citet{King11}.  For CXOU~1229410+075744, the X-ray spectra are well-fitted by disk blackbody models with $\rm{kT}\approx$1 keV \citep{Maccarone11a}. The model of \citet{King11} requires an accretion rate about 11 times the accretion rate required for the source to be at the Eddington luminosity, and hence predicts a characteristic temperature for the source's thermal component of about 0.1-0.2 keV.  For HLX-1, the model of \citet{King11} requires an accretion rate 170 times that needed for the source to be at Eddington, meaning that the thermal component should be in the extreme ultraviolet, undetectable by X-ray observatories, rather than at about 0.2 keV as observed \citep[e.g.][]{Farrell09}.

\section{Conclusions}
\label{sec:conclusions}

Our survey of 174 massive GCs in NGC~4472 identified only one cluster with [OIII] emission. As discussed above, this source is thought to be associated with the LMXB in this cluster, which is accreting at super-Eddington rates. The high X-ray luminosity of this source, combined with the absence of any similar [OIII] sources in the other GCs, favors the theory that this emission is from a BH LMXB, rather than a NS LMXB that is geometrically beamed. The absence of any narrow [OIII] lines in these data also sets constraints on the formation of PNe in these clusters. This limit requires that GCs have some of the lowest PNe formation rates observed. However, galaxies with similarly low rates \textit{are} observed. We note that correlations, observed in elliptical galaxies between their PN population and the galaxies mass and metallicity, do not extend to the low mass and low metallicity region occupied by GCs. Potentially, this may point to other parameters, such as age, being the cause of these correlations. The formation rate of PNe in GCs is significantly lower than the number of post-AGB stars predicted from stellar evolution, suggesting that some of these stars do not produce PNe. The lower rates observed are consistent with the lower rates predicted, if PNe require a binary origin. However, the low $\alpha$ in both the MW's and NGC~4472's cluster systems, suggests that the formation of PNe is not enhanced in GCs due to dynamical formation of tight binaries. Indeed, dynamical interactions may act to reduce the number of PNe, by enhancing mass loss on the RGB/ early ABG branch. Larger samples of GCs with [OIII] data should be highly beneficial to better constrain these conclusions in the future.

\section*{Acknowledgements}

We would like to thank the anonymous referee for their comments, which were beneficial to the final version of this paper. TJM would like to thank Laura Chomiuk for interesting discussions related to the GC PN population. We would like to thank Nelson Caldwell and Jay Strader for discussions related to the M31 GC [OIII] source. MBP would like to acknowledge support from the NASA grant NNX08AJ60G. SEZ would like to acknowledge support from the NSF grant AST-0807557. This paper is based on observations made with ESO Telescopes at the La Silla or Paranal Observatories under programme ID 071.B-0537(A) and 076.B-0446(A).


\label{lastpage}

\end{document}